\documentclass{aa} 
\usepackage{times,amssymb,amsmath} 
\usepackage{epsfig} 
\usepackage{lscape}

\newcommand{\mrm}{\mathrm}

\begin{document} 

\title{Evolution in the properties of Lyman-$\alpha$ emitters from redshifts $z \sim 3$ to $z \sim 2$}

\author{K.K. Nilsson\inst{1}
        \and C. Tapken\inst{1}
        \and P. M\o ller\inst{2}
        \and W. Freudling\inst{2}
        \and J.P.U. Fynbo\inst{3}
        \and K. Meisenheimer\inst{1}
        \and P. Laursen\inst{3}
        \and G. {\"O}stlin\inst{4}
}

\institute{
   Max-Planck-Institut f{\"u}r Astronomie, K{\"o}nigstuhl 17,
   69117 Heidelberg, Germany\\	
\and
   European Southern Observatory, Karl-Schwarzschild-Stra\ss e 2, 85748
   Garching bei M\"unchen, Germany\\
\and
   Dark Cosmology Centre, Niels Bohr Institute, University of Copenhagen, 
   Juliane Maries Vej 30, 2100 Copenhagen $\O$, Denmark\\
\and
   Stockholm Observatory, Department of Astronomy, Stockholm University, 
   AlbaNova University Centre, 106 91 Stockholm, Sweden\\
}
\offprints{knilsson@mpia-hd.mpg.de}
\date{Received date / Accepted date}
\titlerunning{Ly$\alpha$ emitters at $z = 2.25$}

\abstract{
Narrow-band surveys to detect Ly$\alpha$ emitters are powerful tools for identifying 
high, and very high, redshift galaxies. Although samples are increasing at
redshifts $z = 3 - 6$, the nature of these galaxies is still poorly known. 
The number of galaxies detected at redshifts below $z \sim 3$ are also small.}
{
We study the properties of $z = 2.25$ Ly$\alpha$ emitters and compare them with those of
$z > 3$ Ly$\alpha$ emitters.}
{
We present narrow-band imaging made with the MPG/ESO 2.2m telescope and the
WFI (Wide Field Imager) detector. Using this data, we have searched
for emission-line objects. We find 170 candidate typical Ly$\alpha$ emitters and 17 candidates that we regard as high UV-transmission Ly$\alpha$ emitters. 
We have derived the magnitudes of these objects in 8 photometric bands from $u^*$ to $K_s$,
and studied whether they have X-ray and/or radio counterparts.   }
{
We demonstrate that there has been significant evolution in the properties of Ly$\alpha$ emitters between
redshift $z \sim 3$ and $z = 2.25$. The spread in spectral energy distributions (SEDs) at the lower redshift is larger and we detect a significant AGN contribution in the sample. The 
distribution of 
the equivalent widths is narrower than at $z \sim 3$, with only a few candidates with rest-frame equivalent width above the
predicted limit of 240~{\AA}. The star formation rates derived from the Ly$\alpha$ emission compared to those derived from the UV emission are lower by on average a factor of $\sim 1.8$, indicative of a significant absorption by dust. }
{
Ly$\alpha$ emitters at redshift $z = 2.25$ may be more evolved than Ly$\alpha$ emitters at higher redshift. The red SEDs imply more massive, older and/or dustier galaxies at lower redshift than observed at higher redshifts. The decrease in equivalent widths and star formation rates indicate more quiescent galaxies, with in general less star formation than in higher redshift galaxies. At $z = 2.25$, AGN appear to be more abundant and also to contribute more to the Ly$\alpha$ emitting population. }

\maketitle

\section{Introduction}
Over the past decade, a large number of so-called Lyman-$\alpha$ (Ly$\alpha$) 
emitters have been discovered at high redshift. Several techniques have been
employed in the search for these galaxies, but the by far most common method
is that of narrow-band imaging, where a narrow-band filter is tuned to 
Ly$\alpha$ within a particular narrow redshift range. Objects with large equivalent
widths (EWs) are thus selected by comparing the colours in the narrow-band image and
complementary broad-band images. Spectroscopically confirmed Ly$\alpha$ emitters now include several hundreds of sources
at redshifts $z \sim 3$ (e.g., M{\o}ller \& Warren 1993; Cowie \& Hu 1998; Steidel et al. 2000; 
Fynbo et al. 2001, 2003; Matsuda 
et al. 2005; Venemans et al. 2007; Nilsson et al. 2007; Ouchi et al. 2008), 
$z \sim 4.5$ 
(Finkelstein et al. 2007), 
$z \sim 5.7$ (Malhotra et al. 2005; Shimasaku et al. 2006; Tapken et al. 2006)
and $z \sim 6.5$ (Taniguchi et al. 2005; Kashikawa et al. 2006). However, in
the low redshift range, between $z \approx 1.6$ corresponding to the
atmospheric cut-off in the UV and $z \sim 3$,
little progress has been made. In practice, the lower redshift limit is
in fact higher than $z \approx 1.6$ because of the drop-off in CCD sensitivity
and a more typical lower redshift limit is $z \sim 2.0$. Eight narrow-band 
surveys have so far been published below $z \sim 3$; Fynbo et al. (1999), Pentericci et al. (2000), Stiavelli et al. (2001), Fynbo et al. (2002, 2003a, 2003b), Francis et al. (2004) and Venemans et al. (2007). Furthermore, Van Breukelen et al. (2005) published a sample of Ly$\alpha$ emitters (LAEs) at 
$z \sim 2.5$ using integral-field spectroscopy. The difficulty in observing
the Ly$\alpha$ line between redshifts $2 < z < 3$ lies in the low throughput of optical
systems, the low efficiency of CCDs in this wavelength range, and higher extinction 
in the atmosphere. Even so, the
advantage of a smaller luminosity distance is rewarding in that a higher 
\emph{flux} limit equals the same \emph{luminosity} limit as surveys at higher
redshift. It also facilitates follow-up observations into the nature of these
objects. 

At higher redshifts, Ly$\alpha$ emitters have been observed to be increasingly bluer, younger and smaller with increasing redshift. At $z \sim 3$, Gawiser et al. (2006) inferred stellar masses of a few~$\times 10^8$~M$_{\odot}$, almost no dust extinction, and ages of the order of $100$~Myr. In a follow-up paper, Gawiser et al. (2007) studied a stacked sample of 52 Ly$\alpha$ emitters without Spitzer detections and confirmed their results of young ages and that no dust appears to be present in these systems, although they inferred slightly higher masses. The galaxies in their sample with Spitzer detections are presented in Lai et al. (2008). For these galaxies, older ages and higher masses are reported and the authors propose that $z \sim 3$ Ly$\alpha$ emitters may have a wide range of properties. In Nilsson et al. (2007), a stacked sample of Ly$\alpha$ emitters at $z = 3.15$ were studied. Here, small masses and low dust contents were inferred, but the ages were unconstrained. At even higher redshift, Pirzkal et al. (2007) showed that a sample of $4 < z < 5.7$ Ly$\alpha$ emitters had very young ages of a few Myr and small stellar masses, in the range $10^6 - 10^8$~M$_{\odot}$. These results agreed well with those of Finkelstein et al. (2007), who studied almost 100 Ly$\alpha$ emitters at $z = 4.5$. Finkelstein et al. (2008) reported, studying a different sample, finding very dusty, massive galaxies with $A_{1200}$ as high as 4.5 magnitudes and masses as high as several $\times 10^{10}$~M$_{\odot}$, at $z = 4.5$. They argued that they observed a bimodality in the properties of Ly$\alpha$ emitters; young and blue galaxies versus old and dusty. Thus, the properties of $z \geq 3$ Ly$\alpha$ emitters are poorly constrained, but these galaxies tend to be considered young, blue, small and dust-free. With the data presented here, we aim to extend the study of the properties of Ly$\alpha$ emitters to lower redshifts, where a wider range of the SED can be studied in the optical/infrared and the luminosity distance is smaller, allowing a more detailed analysis. 

This paper is organised as follows. Section~\ref{sec:obs} presents the observations leading to our sample, as well as the data reduction. Section~\ref{sec:sel} includes the object detection and candidate selection and a discussion about possible interlopers in the sample. In Sect.~\ref{sec:basic} we present all the basic characteristics of the sample, including photometry, AGN contribution, surface density, and sizes, and the equivalent width distribution of the candidates. We summarise the results in Sect.~\ref{sec:disc}.

\vskip 5mm
Throughout this paper, we assume a cosmology with $H_0=72$
km s$^{-1}$ Mpc$^{-1}$ (Freedman et al. 2001), $\Omega _{\rm m}=0.3$ and
$\Omega _\Lambda=0.7$. Magnitudes are given in the AB system. 

\section{Observations}\label{sec:obs}
\begin{figure}[t]
\begin{center}
\epsfig{file=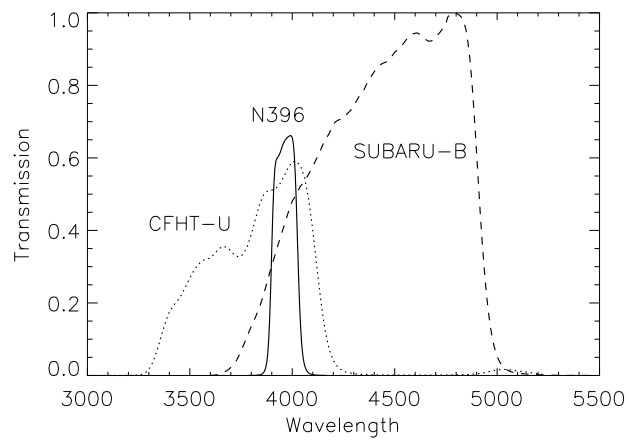,width=9.0cm}
\caption{Transmission of selection filters. The WFI narrow-band
filter is drawn with a solid line. The dashed line shows the SUBARU Bj band 
filter curve and the dotted line the CFHT \emph{u$^*$} band filter curve.}
\label{filtercurve}
\end{center}
\end{figure}

In March 2007, a 35$\times$34 arcmin$^2$ section of the COSMOS field, centred on
R.A.~$ = 10^h 00^m 27^s$  and Dec~$ = 02^{\circ} 12' 22$\farcs$7$ (J2000),
was observed with the Wide-Field Imager (WFI; Baade et al. 1999) on the MPG/ESO 2.2m telescope
on La Silla. A log of the observations can be found in Table~\ref{obslog}.
\begin{table}[t]
\begin{center}
\caption{Log of imaging observations with WFI. }
\begin{tabular}{@{}lcccccc}
\hline
\hline
Date & Total exp. & Average seeing  & \\
\hline
10-11.03.2007   &   0.56 hours   &   1\farcs40  &  \\
11-12.03.2007   &   0.83 hours   &   1\farcs24  &  \\
12-13.03.2007   &   1.33 hours   &   1\farcs04  &  \\
13-14.03.2007   &   2.50 hours   &   1\farcs65  &  \\
14-15.03.2007   &   1.67 hours   &   1\farcs26  &  \\
15-16.03.2007   &   1.45 hours   &   1\farcs81  &  \\
16-17.03.2007   &   3.33 hours   &   1\farcs11  &  \\
17-18.03.2007   &   1.67 hours   &   1\farcs32  &  \\
18-19.03.2007   &   3.33 hours   &   1\farcs15  &  \\
19-20.03.2007   &   2.50 hours   &   1\farcs25  &  \\
20-21.03.2007   &   2.50 hours   &   1\farcs36  &  \\
22-23.03.2007   &   1.67 hours   &   1\farcs13  &  \\
\hline
\label{obslog}
\end{tabular}
\end{center}
\end{table}
The total dithered image consisted of 29 exposures with a total exposure time
of 99624 seconds, or 27.7 hours. The observations were made with narrow-band 
filter N396/12 with a central wavelength of 396.3 nm and a FWHM of 12.9 nm. 
This wavelength range corresponds to $z = 2.206 - 2.312$ for Ly$\alpha$, $z = 0.046 - 0.081$ for 
[OII], and $z = 1.52 - 1.59$ for CIV and the surveyed comoving volumes 
(after masking, see Sect.~\ref{selection}) are 
$\sim 329$~$300$~Mpc$^{-3}$, $820$~Mpc$^{-3}$ and $\sim 225$~$000$~Mpc$^{-3}$, respectively.
The narrow-band filter curve is shown with the filter curves of the filters used
for the selection of candidates in Fig.~\ref{filtercurve}.

The data were reduced using the \emph{MPIAphot} pipeline developed in 
MIDAS with
specific routines to handle WFI data. We briefly describe the 
reduction steps performed on the data. The data from the individual CCDs were converted into MIDAS format and bias-corrected on each separate
CCD. The individual CCD frames were then placed into placeholders in an empty
mosaic, thus creating a full mosaic for each frame including the gaps between
the CCDs. To correct for bad pixels and columns as well as correcting columns
with a constant offset compared to the surrounding pixels, lamp images
with exposure times ranging between $1 - 220$~s~were downloaded and analysed.
The science frames were then corrected for both the bad columns and the
offset columns. The frames were then flat-field corrected using a master
flat created from sky flats taken at the time of observations. To determine the offsets between 
the images, an SDSS catalogue of known sources was used, combined with an 
algorithm that identifies sources and matches them with catalogue entries. The information
about the shift is then entered into the header. In the final step, the images were shifted 
and rebinned onto a gnomonic projection (i.e. by de-projecting all great circles onto straight lines), 
cosmic-ray hits were removed and a final, co-added image
created. Significant residuals from the background subtraction were apparent in the
mosaiced image, in particular around the edges of the individual images used
to create the mosaic. We used the following procedure to remove it. First, we used
the ``clean'' option of the IRAF task \emph{imsurfit} to create a version of the image
in which  objects had been removed and interpolated over. We then edited this image
interactively to remove any residual flux from objects.  Subsequently, we
fitted a 20$\times$20 piece bicubic spline to this image. We then
subtracted this fit from the original image. This procedure reduced the  background
variations to a small fraction of the background noise. 
We then flux calibrated the narrow-band image by calculating the fluxes of all narrow-band selected objects (see Sect.~\ref{selection}) in the CFHT $u^*$ and the SUBARU Bj images (see Table~\ref{cosmodata}) and interpolated the fluxes in the narrow-band image. We can then calculate the zero-point of the image, assuming that the median equivalent width of all objects is zero. The $5 \sigma$ detection limit in a $3''$ diameter aperture in the image is
$25.3$ AB magnitude and the $90$\% completeness limit is $25.1$ AB magnitude, corresponding to Ly$\alpha$ luminosities of $\log{L} = 42.36$~erg~s$^{-1}$ and $\log{L} = 42.44$~erg~s$^{-1}$, respectively, at $z = 2.25$.

\section{Selection of candidates}\label{sec:sel}
\subsection{Object detection and candidate selection}\label{selection}

For object detection, we used the SExtractor software (Bertin \& Arnouts 1996).
The narrow-band image was used as a detection image, and objects with 
a minimum of 8 adjoining pixels and a threshold of $2\sigma$ per pixel were selected. 
For the selection, we used 
two broad-band filters corresponding to wavelengths to both the blue and red side of the narrow-band 
filter, the CFHT $u^*$ band image and the SUBARU Bj band image, see also 
Fig.~\ref{filtercurve}. Both images were taken from the public data in the COSMOS field 
(Capak et al. 2007). The $u^*$ and Bj band images were created from combining several
smaller sub-images to match perfectly the field of the narrow-band image, and
they were subsequently re-binned and smoothed to match the pixel size and 
source PSF of the narrow-band image. The flux of each object detected in
the narrow-band image was then measured in all three bands in circular apertures 
with a diameter of
$3''$, where each aperture was centred on the position of the object
detected in the narrow-band image. For the catalogue, we included only
objects with $S/N > 5$ in the narrow-band image that were found at least $70$~pixels from the 
edge of the image. We also masked areas where stray light and bright stars affected the image. 
This left us with a catalogue of $21$~$275$~objects in an effective area of $\sim 1014$~arcmin$^2$.

\begin{figure}[!t]
\begin{center}
\epsfig{file=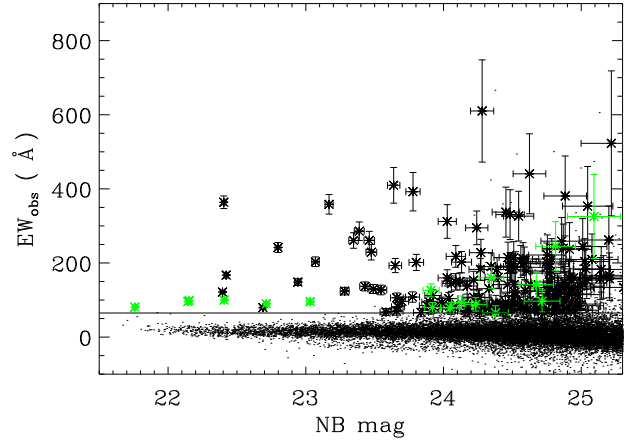,width=9.2cm}
\caption{Plot of the selection criteria, with narrow-band magnitude on the $x$-axis and observed equivalent width on the $y$-axis. Selected candidates are shown with error bars, green points correspond to the GALEX-detected objects, see sec.~\ref{interloper}. Solid line marks the 65~{\AA} limit. Objects above this line with EW significance greater than $2.3 \sigma$ are selected in the first catalogue. Dots mark the entire sample of $5\sigma$ detections in the narrow-band, and dots above the selection criteria without error bars were rejected in the visual inspection.}
\label{selplot}
\end{center}
\end{figure}

\begin{figure}[!t]
\begin{center}
\epsfig{file=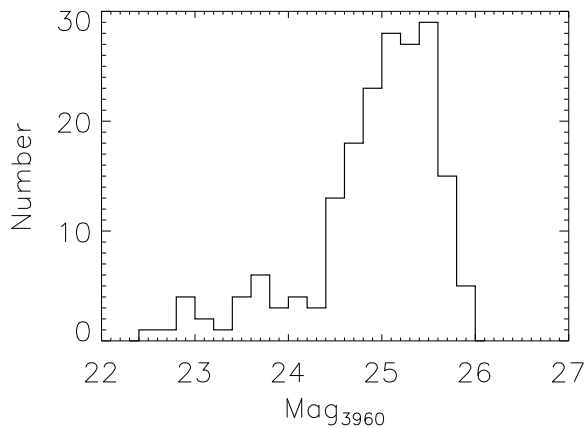,width=9.2cm}
\caption{Histogram of Ly$\alpha$ magnitudes (continuum-subtracted narrow-band magnitudes), Mag$_{3960}$, of emission-line candidate sample. Binsize is 0.2.}
\label{fig:nbmag}
\end{center}
\end{figure}

The candidate Ly$\alpha$ emitters were selected according to a similar method
as described in Nilsson et al.~(2007). The method is 
based on determining the
equivalent width (EW) of a potential emission/absorption line located in the
narrow-band wavelength range. Thus, the flux density was determined in the CFHT
$u^*$ and SUBARU Bj bands and interpolated between the central wavelengths of these
filters and the central wavelength of the narrow-band filter. The EW was then 
determined by dividing the measured flux in the narrow-band with the interpolated flux density.
Simultaneously, we also calculated the propagated statistical error in this
measurement. The calculated EW represents a lower limit to the true EW of the
object as there may be several objects, unrelated to the narrow-band object, 
detected inside the aperture in the broad-band images. The EW calculation is also complicated by
the uncertainties in the continuum interpolation (Hayes \& {\"O}stlin 2006; Hayes et al 2008). To compile the final catalogue
of Ly$\alpha$ emitting candidates, we used two selection criteria. First, all objects with 
a measured EW larger than or equal to $65$~{\AA} (corresponding to $20$~{\AA} in the 
restframe for Ly$\alpha$) are selected. To further exclude false candidates at the faint magnitude end of the sample, in which the EW measurement is dominated by the broad-band flux uncertainties, a significance of the EW greater than 2.3 is required. These criteria are designed 
to detect all emission-line objects with large equivalent widths accurately. This final 
catalogue includes $386$ objects. Following the procedure of Fynbo et al. (2002), three co-authors then individually ranked these $386$ candidates by visual inspection into three categories; first a category of rejected detections (caused by CCD or stellar artifacts or enhancement due to airplane/satellite tracks), then into two categories of likely candidates and candidates that are unlikely to be real but for which no obvious reason was found to reject them (see also Fig.~\ref{selplot}).
The $386$ candidates were split into the different categories of [Likely, Unlikely, Rejected] = [187, 61, 138]. In Table~\ref{ewtab}, we list the coordinates, EWs, significance of the EWs, continuum-subtracted narrow-band magnitudes Mag$_{3960}$ and mean UV spectral index $\beta$ (see Sect.~\ref{sec:phot}) 
of all likely candidates (coordinates of unlikely candidates may be obtained from the authors). 
In the following sections, we only include the candidates in the likely category, which provides a total of 187 candidates.
\begin{table*}[t]
\begin{center}
\caption{Coordinates, observed equivalent widths, continuum subtracted narrow-band magnitudes and UV spectral index $\beta$ of the candidates. Full table is available in the online version.}
\begin{tabular}{@{}lrccccccccc}
\hline
\hline
\multicolumn{8}{c}{Rank~1 candidates} \\
\hline
& LAE\_COSMOS\_\# & R.A. & Dec. & EW (\AA) &  Mag$_{3960}$ & $\beta$ & & \\
\hline
& 1 & 150.05310 & 1.94797 & $142.7\pm38.6$ & $25.50\pm0.34$ & $-0.48\pm0.24$ & & \\
& 2 & 150.21068 & 1.95100 & $225.8\pm57.0$ & $25.41\pm0.32$ & $-0.82\pm0.33$ & & \\
& 3 & 150.29437 & 1.95107 & $564.0\pm43.2$ & $24.24\pm0.09$ & $-1.40\pm0.40$ & & \\
& 4 & 150.25458 & 1.95224 & $118.4\pm33.4$   & $25.55\pm0.36$ & $-2.12\pm0.33$ & & \\
& 5 & 150.31830 & 1.95244 & $279.3\pm41.8$ & $24.91\pm0.18$ & $-2.08\pm0.44$ & & \\
& 6 & 150.37505 & 1.95619 & $95.9\pm24.2$      & $25.46\pm0.32$ & $-1.84\pm0.29$ & & \\
& 7 & 150.15936 & 1.95717 & $790.3\pm140.1$ & $25.04\pm0.21$ & $-2.97\pm1.06$ & & \\
& 8 & 150.37990 & 1.96143 & $119.6\pm16.8$ & $24.88\pm0.16$ & $1.81\pm0.22$ & G & \\
& 9 & 150.28378 & 1.96167 & $248.1\pm50.2$   & $25.20\pm0.25$ & $-1.84\pm0.34$ & & \\
& 10 & 150.30785 & 1.96243 & $81.0\pm28.5$     & $25.80\pm0.47$ & $0.54\pm0.26$ & & \\
\hline
\label{ewtab}
\end{tabular}
\end{center}
\begin{list}{}{}{}{}{}{}
\item[$^{\mathrm{a}}$] Coordinates are in J2000. Candidates marked with the capital letter G are detected with GALEX, see text below. 
\end{list}
\end{table*}
The Ly$\alpha$ magnitudes of our LAE candidates are given by the range Mag$_{3960} = 22.57 - 25.88$ corresponding to Ly$\alpha$ luminosities in the range of $\log{L_{\mathrm{Ly\alpha}}} = 42.13 - 43.45$ (see also the histogram of magnitudes in Fig.~\ref{fig:nbmag}). 

\subsection{Interloper discussion}\label{interloper}

The only possible contaminants in the sample are [OII] emitters at $z = 0.06$ and CIV emitters at 
$z = 1.56$. For [OII] emitters at 
this redshift, the detection limit corresponds to a luminosity limit of $\log{L} = 38.7$~erg~s$^{-1}$ and the 
survey volume is 820~Mpc$^{-3}$. For CIV emitters, the corresponding limit is $\log{L} = 42.0$~erg~s$^{-1}$ and the survey volume is $\sim 225$~$000$~Mpc$^{-3}$. Unfortunately, no published luminosity function for local Universe
[OII] emitters or CIV emitters reach these faint luminosity limits, but for [OII] emitters, extrapolating from the luminosity functions of 
Hogg et al. (1998), Gallego et al. (2002), and Teplitz et al. (2003), we conclude that we could expect
between a few to 200 [OII] emitters in the survey volume. To constrain this value further, the [OII] luminosity function was extrapolated from the H$\alpha$ luminosity function of Ly et al. (2007). Assuming a conversion between [OII] and H$\alpha$ luminosities given by the star formation rate equations of Kennicutt (1998) and Kewley et al. (2004), the [OII] luminosity limit corresponds to an H$\alpha$ luminosity limit of $\log{L} = 38.8$~erg~s$^{-1}$ in this survey. To this limit, Ly et al. (2007) detected $\sim 0.2$ emitters per Mpc$^{-3}$, corresponding to approximately 160 objects in our survey volume, of which only 2\% are expected to have an EW larger than the selection criteria limit of EW$_{\mrm{obs}} = 65$~{\AA} (Hogg et al. 1998). Based on the results of Ly et al. (2007), approximately three [OII] emitters are hence included in the list of candidates. 

We also extracted the photometric redshifts for our candidates from the catalogue of Gabasch et al. (2008). The catalogue is incomplete in areal coverage and covers roughly 80\% of our field. The catalogue was searched for counterpart objects within a $2''$ radii from the position of our source. If several objects were found, the object nearest to our source was chosen as the counterpart. For a total of 187 candidates, 132 had detected counterparts. Unfortunately, the near-IR coverage of the COSMOS field is patchy and shallow, which complicates any photometric redshift determination in the redshift range $z \sim 1.3 - 2.5$, since in this range the Lyman-break has not yet entered the optical window and the Balmer break is in the near-IR. This means that we are in principle able to exclude [OII] emitters from the sample, whereas other intermediate redshift emitters, such as CIV, are difficult to exclude. The redshifts of our 132 detected candidates also show that no emitters are consistent with being [OII] emitters, and the median redshift of the candidates is $z = 1.7$, with most of the candidates being consistent with having a redshift of $z = 2.25$ to within $3\sigma$ margins of error. To study how many CIV emitters may be expected in the survey volume, we used the AGN luminosity function of Bongiorno et al. (2007). Since CIV emission of restframe equivalent widths larger than $EW_0 \gtrsim 25$ are only expected to originate in AGN, and with the luminosity limits in the optical of this survey, the expected number of CIV emitters in our survey volume is $\sim 2$. Any contamination of CIV emitters in the total sample of candidates is thus minimal. 

The COSMOS field also has publicly available GALEX near- and far-UV data. This data may be used to exclude any remaining [OII] emitters from the sample, since the limiting magnitude of the GALEX data is $\sim 25.5$ in the NUV band and, assuming a flat spectrum, all [OII] emitters in the sample should be detected in the GALEX data based on the flux limit in the Bj band. Ly$\alpha$ emitters may or may not be detected in the GALEX data, depending on the level of absorption from the Lyman forest. Ly$\alpha$ emitters would require a transmission of less than $60$\% to be undetected in the GALEX images. M{\o}ller \& Jakobsen (1990) inferred that approximately 10\% of galaxies at redshift $z = 2.25$ have transmissions in the UV higher than $60$\%, in agreement with recent observations at $z \sim 1.5$ (Siana et al. 2007). The conclusion is that we expect all [OII] emitters to be detected in the GALEX images, whereas the total sample of GALEX-detected objects is a mix of high transmission Ly$\alpha$ emitters and [OII] emitters. Of the 187 candidates, 17 are detected in the GALEX images (labelled by a G in Table~\ref{ewtab}). The largest EW of a GALEX-detected candidate is EW$_{\mrm{obs}} = 604$~{\AA}. As shown earlier, the expectation is $\sim 3$ [OII] emitters in the survey volume, and so most of the GALEX-detected objects are most likely to be high transmission Ly$\alpha$ emitters. We separate the GALEX-detected sample and the non-GALEX-detected sample in the following analysis. Thus, we have two samples; one robust sample of $170$ Ly$\alpha$ emitter candidates (removing 17 GALEX detections from a total of 187 candidates), which is expected to be free of interlopers, and one sample of 17 GALEX-detected objects, possibly including a small contamination of [OII] emitters but consisting mostly of high-transmission Ly$\alpha$ emitters. The non-GALEX-detected sample is a conservatively selected sample defined so that the probability of having non-LAE objects in it is minimal, but at the price that it is less complete. For this reason, it is unsuitable for determining volume and surface densities. In Sect.~\ref{sec:density} we return to this issue and define a sample suited for this purpose. From here on, whenever the  ``sample'' is mentioned it is referring to the non-GALEX-detected sample.

\section{Basic characteristics of LAEs}\label{sec:basic}
\subsection{Photometry}\label{sec:phot}
For the photometry of the candidates, we use the public data-set from the COSMOS survey (Capak et al. 2007). The photometry presented here has been made in the bands found in Table~\ref{cosmodata}.
\begin{table}[t]
\begin{center}
\caption{COSMOS broad-bands used.  }
\begin{tabular}{@{}lcccccc}
\hline
\hline
Band & Observatory  & CWL (\AA) & FWHM (\AA) & Depth  & \\
\hline
\emph{u$^*$}     &   CFHT          &  3798   & 720     & 26.4  & \\
Bj                           &   SUBARU   &   4460  & 897     & 27.3  & \\
\emph{g$^+$}     &  SUBARU    &  4780   & 1265   & 27.0  & \\
Vj                           &   SUBARU   &   5484  & 946     & 26.6  &  \\
\emph{r$^+$}      &   SUBARU   &  6295   & 1382  & 26.8   & \\
\emph{i$^+$}      &   SUBARU   &   7641  & 1497  & 26.2   &  \\
\emph{z$^+$}     &   SUBARU   &   9037  &  856   & 25.2   &  \\
$K_s$                  &   KPNO         &  21500 & 3200 & 21.6   &  \\
\hline
\label{cosmodata}
\end{tabular}
\end{center}
\begin{list}{}{}{}{}{}{}
\item[$^{\mathrm{b}}$] The depths are $5\sigma$ AB magnitudes as measured in $3''$ apertures (Capak et al. 2007).
\end{list}
\end{table}
All images are on the common COSMOS tiling grid and were cut and re-binned to match the narrow-band image in coverage and pixel size. We ran SExtractor in dual image mode, measuring the flux in apertures defined in the narrow-band image, also including the RMS images in the measurement through the WEIGHT\_IMAGE option. The apertures had a diameter of $3''$. Magnitudes were calculated using the zero-point given in the COSMOS data release. These magnitudes will, in the case of the candidates, be underestimated, since objects unrelated to the LAE candidate may be blended into the aperture (see below for a discussion of contamination in the aperture). In the following we exclude the results in the \emph{g$^+$} band as these magnitudes appear to be inconsistent. When plotting the SEDs of the objects, the \emph{g$^+$} band magnitudes are consistently lower than the other SED points by $0.1 - 0.5$~magnitudes. 

\begin{figure}[t]
\begin{center}
\epsfig{file=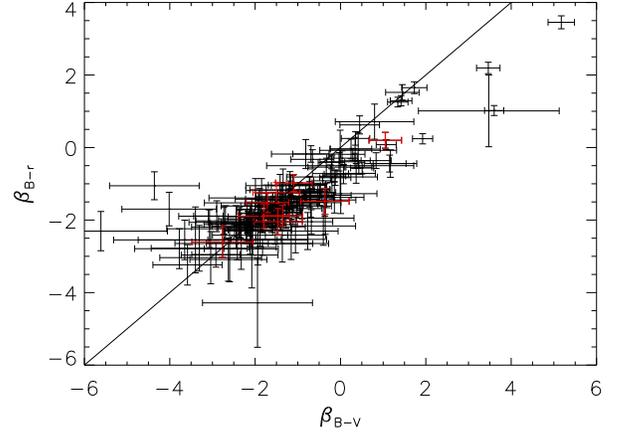,width=9.0cm}
\caption{UV spectral index calculated from the Bj$-$Vj and Bj$-r^+$ colours. The two colours give approximately the same results, as seen by the solid line marking a linear relation. The spread in the spectral index is very large. AGN (see sec.~\ref{sec:agn}) are marked in red. Objects with red SEDs can be found in the upper right corner and objects with blue SEDs in the lower left corner. }
\label{beta}
\end{center}
\end{figure}

We calculated the UV spectral index of $\beta$ for the candidates, which is defined to be:
\begin{equation}
f_\lambda \propto \lambda^{\beta}
\end{equation}
where $f_\lambda$ is the flux density per wavelength interval in the UV range. The spectral index is calculated as in Meurer et al. (1997) and Hathi et al. (2008) for the Bj, Vj, and $r^+$ bands using the following equations:
\begin{equation}\label{eq:betabv}
\beta_{\mrm{Bj}-\mrm{Vj}} = 4.456 \times (\mrm{Bj} - \mrm{Vj}) - 2
\end{equation}
\begin{equation}\label{eq:betabr}
\beta_{\mrm{Bj}-r^+} = 2.673 \times (\mrm{Bj} - r^+) - 2
\end{equation}
where Bj, Vj, and $r^+$ are the AB magnitudes in the respective bands. A plot of the results can be seen in Fig.~\ref{beta} and the total $\beta$, calculated to be the mean of the two measurements as described in Eq.~\ref{eq:betabv} and \ref{eq:betabr}, for our LAE candidates is found in Table~\ref{ewtab}.
Most of the candidates are within the expected range of $-2 < \beta < 0$, with some outliers at both larger and smaller $\beta$. Of the four objects with $\beta_{Bj-Vj} \geq 2$, one is also detected in the $K_s$ images, and two are affected by contamination from nearby objects. These objects are in the mid range of narrow-band magnitude, but in general have higher fluxes in the Vj band. On the other end of the scale, the three objects with $\beta_{Bj-Vj} \leq -4$ are all drawn from the faint end of the magnitude distribution in both the narrow-band and the Vj band, which explains their large error bars. 

\begin{figure}[!t]
\begin{center}
\epsfig{file=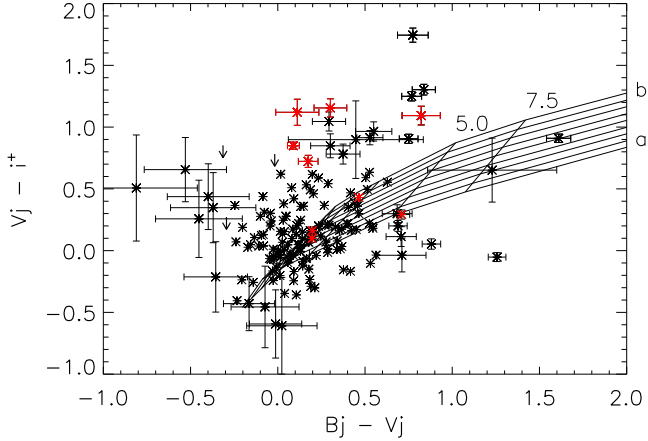,width=9.2cm}
\caption{Colour-colour diagram with Bj$-$Vj on the $x$-axis and Vj$ - i^+$ on the $y$-axis. Objects with upper limits (on the flux) in one band are shown as arrows with 3$\sigma$ upper limits and objects with a colour in either bands that is $1.5\sigma$ away from the median colour in that band are shown with error bars. AGN are marked in red. The solid lines display the evolution of a single stellar population with increasing dust and age; age causes the objects to move predominantly to the right in the plot and dust predominantly upwards. The markers $5.0$ and $7.5$ correspond to age steps of $5.0 \times 10^8$ and $7.5 \times 10^8$~yrs and the dust steps go from $E(B - V)$ of 0.0 (point \emph{a} in the plot) to 1.0 (point \emph{b} in the plot) in steps of 0.125. }
\label{bvicol}
\end{center}
\end{figure}

The SEDs of the candidates have characteristics that are atypical of higher redshift LAE SEDs. It has previously been reported that LAEs have SEDs that are in many cases blue (Gawiser et al. 2006, 2007; Pirzkal et al. 2007). Only two other surveys reported detections of red SEDs in several LAEs (Stiavelli et al. 2001; Lai et al. 2008), although one extremely red object was found by Nilsson et al. (2007). Among the 170 objects detected in this survey, a total of 118 candidates have Vj$-i^+$ colours greater than zero, indicative of red colours, and five candidates have undefined colours due to upper limits in both bands, or only in the $i^+$ band. The median values of Bj$-$Vj and Vj$-i^+$ among candidates with 3$\sigma$ detections in both bands is 0.14 and 0.15, respectively. In Fig.~\ref{bvicol}, the colours Bj$-$Vj and Vj$- i^+$ are plotted in a colour-colour diagram. 
In this figure, the tracks of a single stellar population, with increasing age and dust content, are superimposed. These populations were simulated with GALAXEV (Bruzual \& Charlot 2003) in age steps of 1, 10, 50, 100, 500 and 750 Myrs and 1 and 2 Gyrs. The dust extinction was increased in 9 steps of $E(B - V)$ between 0.0 and 1.0 with increments of 0.125 using the Calzetti extinction law (Calzetti et al. 2000). We also completed simulations of an exponentially decaying star formation rate (SFR) and constant SFR. However, the colours of these populations occupy a smaller region of the diagram than those of the single stellar populations and were omitted from the plot. Most of the candidates are consistent with being young galaxies with ages in the range $10^6 - 10^8$~yrs with various dust contents. Some objects appear evolved with a range of dust content. There are also some objects that appear evolved but fall below even the zero-dust line in the plot, which may be affected by nearby objects. Another group of objects are very red in terms of their Vj$ - i^+$ colours but are too blue in Bj$ - $Vj to be explained by simple stellar populations. These may be reproduced more accurately by models with more complicated star formation histories. 

\begin{figure}[!t]
\begin{center}
\epsfig{file=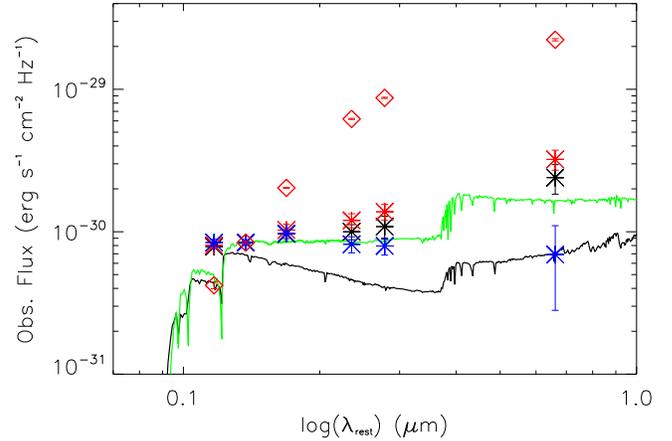,width=9.2cm}
\caption{Stacked SED of the sample. Points are the stacked magnitudes in six bands ($u^*$, Bj, Vj, $i^+$, $z^+$ and $K_s$) and in three bins; black points are the full sample, excluding $K_s$-detected galaxies and galaxies with bright nearby objects, red stars are those with red Vj$-i^+$ colours and blue points those with blue Vj$-i^+$ colours. Red diamonds mark the stacked $K_s$-detected sample without AGN-detected galaxies. All are normalised to the Bj flux of the Nilsson et al. (2007) SED. The lines are two best-fit SEDs of $z \sim 3$ LAEs from Nilsson et al. (2007; green line) and Gawiser et al. (2007; black line). }
\label{fig:redsed}
\end{center}
\end{figure}

In the sample, 12 objects have $K_s$ band detections above the limiting magnitude of 22.2 (3$\sigma$, AB), corresponding to 7\% of the complete sample. Note that four of these 12 $K_s$-detected objects are also selected as AGN (see Sect.~\ref{sec:agn}). The $K_s$-detected objects are all drawn from the brighter end of the candidate sample, but  have a range in Vj magnitude of $21.5 - 26$, whereas the entire candidate sample have a range in magnitude of $21.5 - 27$, i.e. not only the brightest candidates have $K_s$ detections. We stacked the thumb-nail images of the candidates without $K_s$ detections, removing objects from the stack with nearby, unrelated, bright detections. The total stack consisted of 144 candidates (with GALEX detections and objects with nearby, unrelated detections removed) and revealed a mean detection in the sample of magnitude $24.00 \pm 0.29$.  The sample was also divided into two subsamples with Vj$-i^+$ colours Vj$-i^+ \geq 0$ and Vj$-i^+ < 0$, respectively. In the subsample with red Vj$-i^+$ colours, consisting of 96 of the 144 candidates, a clear detection with magnitude $23.65 \pm 0.19$ was made. In the subsample with bluer colours, no detection was made and an upper limit of $24.26$ ($3\sigma$) was derived. In Fig.~\ref{fig:redsed}, we show the stacked magnitudes of the sample divided into four bins; the total sample, the sample with Vj$-i^+ \geq 0$ (called the ``red'' sample), the sample with Vj$-i^+ < 0$ (called the ``blue'' sample), and the $K_s$-detected sample. The total sample consists of 144 candidates, the red subsample consists of 96 candidates, and the blue subsample consists of 48 candidates.
The magnitudes in Fig.~\ref{fig:redsed} are also found in Table~\ref{stackmag}.
\begin{table}[t]
\begin{center}
\caption{Stacked magnitudes for the sample. }
\begin{tabular}{@{}lcccccc}
\hline
\hline
 & Total sample  & Red sample & Blue sample & $K_s$-detected  \\
\hline
\emph{u$^*$}     &   $25.21\pm0.03$   &  $25.16\pm0.05$  & $25.34\pm0.06$  & $24.75\pm0.06$  \\
Bj                           &  $25.14\pm0.03$   &   $25.11\pm0.04$ & $25.34\pm0.05$ & $24.00\pm0.02$  \\
Vj                           &  $24.99\pm0.08$   &  $24.89\pm0.11$  & $25.18\pm0.03$ & $23.04\pm0.01$  \\
\emph{i$^+$}      &   $24.95\pm0.14$  &   $24.72\pm0.13$  & $25.37\pm0.15$ & $21.83\pm0.01$   \\
\emph{z$^+$}     &   $24.86\pm0.22$  &   $24.57\pm0.15$ &  $25.40\pm0.16$  & $21.46\pm0.01$   \\
$K_s$                  &   $24.00\pm0.29$  &  $23.65\pm0.19$ &  $25.55\pm0.98^*$ & $20.44\pm0.02$   \\
\hline
\label{stackmag}
\end{tabular}
\end{center}
\begin{list}{}{}{}{}{}{}
\item[$^{\mathrm{c}}$] Upper limits are $3\sigma$. $K_s$-detected sample is without AGN-detected galaxies. $^*$Blue sample $K_s$ magnitude is calculated from the $K_s$ magnitudes of the total and red samples.
\end{list}
\end{table}
In Fig.~\ref{fig:redsed}, it is seen that the stacked SED of the total sample is redder than those at $z \sim 3$; the points of the non-$K_s$-detected galaxies are in principle consistent with the Nilsson et al. (2007) fit within $2\sigma$ but inconsistent across the spectrum with the Gawiser et al. (2007) fit. The normalised $K_s$ magnitude for the total sample is $25.45\pm0.29$, whereas the same magnitudes measured from the synthetic SEDs of Nilsson et al. (2007) and Gawiser et al. (2007) are $26.00$ and $26.39$, respectively. Hence, roughly two-thirds of the entire sample exhibit on average more red SEDs than at $z \sim 3$. Noteworthy is also that SED of the blue sample (with Vj$-i^+ < 0$) is consistent with the $z \sim 3$ SEDs. The Vj$- K_s$ slopes of the red and blue samples are $1.24 \pm 0.22$ and $-0.37\pm0.98$, indicating a clear difference in the properties of the blue and red samples of galaxies. For reference, the Vj-$K_s$ slope of the $K_s$-detected sample is $2.68\pm0.04$. A large spread in colours is evident in this survey, from the ``classical'' blue, flat SED galaxies to galaxies with very red SEDs, where the redder SEDs constitute two-thirds of the sample (see also Finkelstein et al. 2008 for a similar result). In a future paper, the Spitzer fluxes of these galaxies will be studied. 

To study the level of contamination in the apertures, and the possibility that the very red SEDs originate from this source, the narrow-band catalogue was searched for other detected objects within two aperture radii (i.e. within $3''$) of each LAE candidate. Of the 170 non-GALEX-detected candidates, 26 have one or more detected objects within this search radius. This corresponds to a potential contamination rate of 15~\% in the photometric measurements. Two objects have more than one nearby object. We also wish to determine if the apparent high number of red objects is caused by flux contamination from unrelated objects in the aperture. Of the 26 candidates with nearby objects, 19 have a Vj$ - i^+$ colour in the aperture greater than zero, and 15 have Vj$ - i^+$ colours larger than the median of Vj$ - i^+$ for all 170 candidates. This implies that the contamination is not confined to only apparent red objects but affects all types of SEDs. We are thus confident that the contamination in the photometry is small, and that contamination cannot explain the apparent red colours of the LAE candidates. 

\subsection{AGN contribution}\label{sec:agn}
The COSMOS public data includes a release of X-ray data in three bands, $0.5 - 2.0$, $2.0 - 4.5$, and $4.5 - 10.0$ keV, as observed by the XMM observatory (Hasinger et al. 2007). We used the catalogue of objects of Cappelluti et al. (2007) to measure the AGN fraction of all candidates. The catalogue consists of 1390 entries representing point-like sources and gives the fluxes in the three bands including errors. The catalogue detects objects to limiting fluxes of $7.2 \times 10^{-16}$, $4.0 \times 10^{-15}$, and $9.7 \times 10^{-15}$ erg~s$^{-1}$~cm$^{-2}$, respectively, in the bands $0.5-2.0$, $2.0-4.5$, and $4.5-10.0$~keV with a confidence of roughly $4.5\sigma$ (Hasinger et al. 2007).  This catalogue was searched for objects detected within a radius of $8''$ (i.e. 2 pixels in the XMM images) of the candidates. Fifteen of these potential counterparts were detected, of which five were from the list of GALEX-detected candidates. The X-ray fluxes of these candidates are found in Table~\ref{xraytab}.
\begin{table*}[t]
\begin{center}
\caption{X-ray and radio properties of counterparts of the candidates. }
\begin{tabular}{@{}llcccccccccc}
\hline
\hline
\multicolumn{10}{c}{LAE candidates} \\
\hline
LAE\_& $\mrm{S}_{0.5-2.0}$ & $\mrm{S}_{2.0-4.5}$ & $\mrm{S}_{4.5-10.0}$  & HR1 & HR2 & $\log(L_{0.5-2.0})$ & Offset & $\mrm{S}_{1.4\mathrm{GHz}}$ & Offset  \\
COSMOS\_\# & & & & & & & (X-ray) & & (Radio)  \\
\hline
25  & $129.0\pm1.87$ &  $191.0\pm5.63$    & $133.0\pm7.03$ & $0.19\pm0.01$ & $-0.18\pm0.03$ & $45.67\pm0.02$ & 11.06 & --- & ---  \\
36 & --- & --- & --- & --- & --- & --- & --- & $140\pm27$ & 0.10  \\
57  & --- & --- & --- & --- & --- & --- & --- & $47\pm10$ & 6.17  \\ 
82 & $1.25\pm1.97$ & --- & --- & $<0.52$ & --- &  $43.66\pm0.41$ & 1.07 & --- & ---  \\
101 & --- & --- & --- & --- & --- & --- & --- & $137\pm27$ & 0.06  \\ 
113 & $1.21\pm4.10$ & $9.58\pm29.1$ & --- & $0.78\pm0.61$ & $<0.01$ &  $43.64\pm0.64$ & 10.55 & --- & ---  \\
115  &  $7.20\pm6.00$     &  $16.2\pm21.5$ &               ---         & $0.38\pm0.59$ & $<-0.25$ & $44.42\pm0.78$ & 0.17 & --- & ---  \\ 
119 & $0.92\pm0.26$ & --- & --- & $<0.63$ & --- & $43.55\pm0.14$ & 2.67 & --- & ---  \\ 
121  &  $17.6\pm0.73$ & $31.0\pm2.54$ & $18.3\pm3.52$ & $0.28\pm0.04$  & $-0.26\pm0.11$ & $44.81\pm0.03$ & 6.36 & --- & ---  \\ 
129 & $1.10\pm0.27$ & --- & --- & $<0.57$ & --- & $43.60\pm0.12$ & 1.95 & --- & ---  \\ 
132  & --- & --- & --- & --- & --- & --- & --- & $47\pm10$ & 9.83  \\ 
140  &  $17.5\pm0.72$ & $46.3\pm2.97$ & $44.9\pm4.73$ & $0.45\pm0.03$ & $-0.02\pm0.06$ & $44.80\pm0.02$ & 0.74 & --- & ---  \\ 
150  &  $2.6\pm0.30$   & $8.3\pm1.48$    &     ---                    & $0.52\pm0.07$ & $<0.08$ & $43.98\pm0.05$ & 0.62 & $107\pm30$ & 0.21   \\ 
161 & $2.42\pm0.33$ & --- & --- & $<0.25$ & --- & $43.95\pm0.06$ & 0.21 & --- & ---  \\
\hline
\multicolumn{10}{c}{GALEX-detected candidates} \\
\hline
42  & $4.9\pm2.92$ &          ---                & --- &      $<-0.10$                   & --- & $40.6\pm0.39$ & 0.55 &  --- & ---  \\
108  & $8.3\pm0.60$ & $16.9\pm2.14$  & --- & $0.34\pm0.06$ & $<-0.27$ & $40.83\pm0.03$ & 1.01 & --- & ---  \\
126  & $7.9\pm0.50$ &  $12.0\pm1.79$ & --- & $0.38\pm0.59$ & $<-0.11$ & $40.81\pm0.03$ & 0.45 & --- & ---  \\
155  & $3.0\pm0.34$ &  $5.6\pm1.21$    & --- & $0.30\pm0.10$ & $<0.27$ & $40.39\pm0.06$ & 1.57 & --- & ---  \\
173  & --- & --- & --- & --- & --- & --- & --- & $110\pm25$ & 9.44  \\
\hline
\label{xraytab}
\end{tabular}
\end{center}
\begin{list}{}{}{}{}{}{}
\item[$^{\mathrm{d}}$] Data from the catalogues of Cappelluti et al. (2007) and Schinnerer et al. (2007). X-ray fluxes are given in units of $10^{-15}$~erg~s$^{-1}$~cm$^{-2}$ and radio fluxes in units of $\mu$Jy. \emph{HR1} and \emph{HR2} correspond to the hardness ratios defined in Eq.~\ref{hr1} and~\ref{hr2}. The X-ray luminosity is in erg~s$^{-1}$, assuming the objects are located at redshifts $z = 2.25$~and~$0.06$ for LAE and GALEX-detected candidates respectively. Offsets are distance from centroid of narrow-band detection to X-ray/radio detection in units of arcseconds. It is expected that the objects with offsets larger than a few arcseconds (25, 57, 113, 121, 132 and 173) are unrelated to the candidates due to the large offset and the relatively large probability of random association. 
\end{list}
\end{table*}
Note that the probability of finding an X-ray object within an aperture of radius $8''$ is $\sim 0.009$, resulting in potentially two random detections in the combined GALEX and non-GALEX-detected sample. For the objects with detections in two or all bands, the hardness ratios can be calculated. Hardness ratios are defined to equal:
\begin{equation}\label{hr1}
\mrm{HR1} = \left(\mrm{S}_{2.0-4.5} - \mrm{S}_{0.5-2.0}\right) / \left(\mrm{S}_{2.0-4.5} + \mrm{S}_{0.5-2.0}\right) 
\end{equation}
\begin{equation}\label{hr2}
\mrm{HR2} = \left(\mrm{S}_{4.5-10.0} - \mrm{S}_{2.0-4.5}\right) / \left(\mrm{S}_{4.5-10.0} + \mrm{S}_{2.0-4.5}\right) 
\end{equation}
The hardness ratios, when applicable, can be found in Table~\ref{xraytab}. All measured hardness ratios (HR1) are positive, indicating that the LAE candidates are type 2 AGN (e.g. Norman et al. 2004). The low luminosity of the GALEX-detected candidates, assuming that they are [OII] emitters indicate that they are normal, star forming galaxies, or that they are Ly$\alpha$ emitters. The total sample of GALEX-detected and non-detected candidates without individual X-ray detections were stacked, and produced no detection in any band in a stack of 170 objects (the nine individual detections were removed from the stack, as well as eight candidates with unrelated objects detected in the vicinity). The non-detection in a stack of 170 objects imply an upper limit to the mean X-ray flux of approximately a factor of ten smaller than those in the catalogue of Cappelluti et al. (2007).

We also searched the catalogue of Schinnerer et al. (2007) for radio counterparts to the candidates. The data on which the catalogue is based was taken with the VLA array and covers the entire COSMOS field to a depth of a few~$\times 10 \mu$Jy. The catalogue presented in Schinnerer et al. (2007) contains 3643 source detections. This catalogue was searched with the same criteria as for the X-ray catalogue (i.e. detections within $8''$ of the candidate positions) and six objects were found. The number of random detections in a $8''$ radius aperture is $0.026$. Of the six radio detections, one object is GALEX-detected (candidate 173). The radio fluxes of these sources can be found in Table~\ref{xraytab}. Only one source has a detection in both the X-ray and radio data. This is in principle not a problem since AGN can be both radio-loud and radio-quiet. The radio objects without X-ray detections are more enigmatic, but can be explained by the relatively shallow luminosity limit of the X-ray observations. If the radio fluxes are converted into star formation rates using the conversion rate of Condon (1992), they correspond to values in the range of $\sim 400 - \sim 1300$~M$_{\odot}$~yr$^{-1}$ whereas the upper limit to star formation rates in the XMM survey is $\sim 1200$~M$_{\odot}$~yr$^{-1}$, using the conversion of Ranalli, Comastri \& Setti (2003).

The AGN fraction of the non-GALEX-detected LAE sample is thus at least $\sim 5$\%, if it is assumed that the X-ray and radio objects within $5''$ of the narrow-band sources are true detections. It is clear that the shallow flux limit to the deepest X-ray band, corresponding to a luminosity of $log(L_{0.5-2.0}) = 43.4$~erg~s$^{-1}$, ensure many AGN of lower luminosities to be undetected, as confirmed by the radio detections without X-ray counterparts. The AGN fraction should therefore almost certainly be even higher, although we did not detect any X-ray flux in the mean, stacked total GALEX and non-GALEX-detected sample of 170 objects. Previous studies of LAEs at higher redshifts exhibited very small AGN contributions, from less than one percent in the $z \sim 4.5$ LAEs studied by Wang et al. (2004) to $\sim 1$~\% at $z \sim 3$ as inferred by Gawiser et al. (2007). Lehmer et al. (2008) calculated the ``AGN fraction function''  of Ly$\alpha$ emitters at $z \sim 3$ both in the field and the overdense region of SSA22 (Fig.~3b in their paper). If we convert our $0.5-2.0$~keV detection limit to their $8-32$~keV luminosity range, we should expect to detect an AGN fraction of $0.5-1.5$~\%, where the lower number is related to their field result and the higher to the SSA22 result. Based on this comprehensive study of X-ray detected Ly$\alpha$ emitting AGN, it is clear that the AGN fraction detected here is larger than at $z\sim3$, as expected from the Lehmer et al. (2008) survey. Ouchi et al. (2008) argued that the missed fraction in AGN X-ray searches among LAEs may be as high as 10\%. We performed the same test as in the Ouchi et al. (2008) publication. Based on the quasar SEDs of Elvis et al. (1994) and the UV spectrum of quasars of Telfer et al. (2002) and Richards et al. (2003) we inferred the ratio of observed $0.5 - 2.0$~keV flux to Ly$\alpha$ flux of $1.87$ and $1.51$ for radio-quiet and radio-loud quasars, respectively . For the flux limit in the X-ray band, the maximum magnitude observed for Ly$\alpha$ is then 23.34. In our complete GALEX and non-GALEX-detected samples, nine objects have magnitudes brighter than this, four of which are GALEX-detected. Of the nine objects, six have X-ray detections (all four GALEX-detected sources and two LAE candidates). Thus, if we exclude the four GALEX-detected sources, two out of five Ly$\alpha$ bright galaxies host AGN and the fraction of X-ray detected AGN in the non-GALEX-detected sample is 40\%, four times the fraction found by Ouchi et al. (2008) for $z\sim3$ LAEs.  To conclude, our detection of a 5\% AGN contribution in the non-GALEX-detected LAE sample is consistent with previous results, but is indicative of a higher AGN fraction being present at this redshift.

\subsection{Surface density, sizes and SFR}\label{sec:density}
For the surface density determination, we included both the GALEX-detected and non-detected objects to obtain as complete as possible a measure of the number density. Since our image has areas of poorer signal-to-noise, as well as stellar artifacts, we selected sub-images of superior quality with a total area of $\sim 288$~arcmin$^2$, corresponding to $\sim 28$\% of the total surveyed area, to use for the surface and volume density calculations. In these areas, the selection is complete, and we found 54 candidates within these sub-images. 
This implies a surface density of LAE candidates of $0.19$~arcmin$^{-2}$~$\Delta z^{-1}$, or $1.91$~arcmin$^{-2}$~$z^{-1}$. The volume density is $0.00062$~Mpc$^{-3}$, which is in the lower range of that observed at redshift $z \sim 3$. Fynbo et al. (2001) summarise the surface densities of several early $z \sim 3$ surveys. With the exception of the Steidel et al. (2000) survey in the overdense SSA22 field, these surveys all determined values of $2.11 - 5.9$~arcmin$^{-2}$~$z^{-1}$ to the flux limit of this survey, but with large error bars. Hence, the values at $z \gtrsim 3$ are consistently higher than our value, but in most cases the measurements agree to within $1\sigma$. For LAE candidates brighter than the  ($5\sigma$) luminosity limit of this survey, Nilsson et al. (2007) identified six candidates in their survey, corresponding to $0.0018$~Mpc$^{-3}$, which represents a decrease of roughly a factor of three although based on a small sample. Gronwall et al. (2007) determined a space density of $0.00057$~Mpc$^{-3}$ above the $5\sigma$ luminosity limit of our survey, which is consistent with the space density found here. Stiavelli et al. (2001) presented a survey of $z = 2.4$ LAEs and found a volume density of $0.0001$~Mpc$^{-3}$ above a luminosity limit of $\log{L} = 42.93$~erg~s$^{-1}$. The same number for this survey is $0.00009$~Mpc$^{-3}$. The numbers agree reasonably well with previous results. Finally, Prescott et al. (2008) completed a survey of $z \sim 2.7$ LAEs around a so-called Ly$\alpha$ blob (e.g. Steidel et al. 2000; Matsuda et al. 2004; Dey et al. 2005; Nilsson et al. 2006). They argued that the central part of their field is overdense, and determined a number density of $0.0021$~Mpc$^{-3}$. This is roughly three and half times higher than in our survey. At the edge of their field, the density is instead $0.0012$~Mpc$^{-3}$, in closer agreement with, but still higher than, our result. Thus, the surface density in this survey is in almost all cases lower than in higher redshift surveys, but could also be consistent with previous results. If one considers that number densities have been found to vary by factors of $2 - 5$ in $0.2$~deg$^2$ fields at redshift $z = 3$ (Ouchi et al. 2008), it is clear that more data at both higher and lower redshifts is needed to resolve this issue finally.

\begin{figure*}[!t]
\begin{center}
\epsfig{file=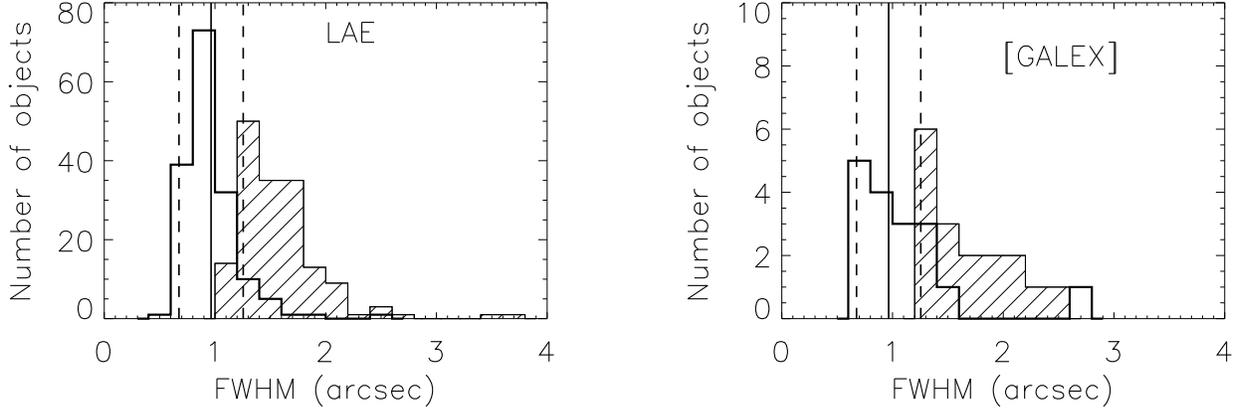,width=18.0cm}
\caption{Size distribution of candidates as measured in the narrow-band (hatched histogram) and $r^+$ (empty histogram) images for LAE candidates and GALEX-detected candidates respectively. Sizes as measured by the FLUX\_RADIUS command in SExtractor. The solid line represents the PSF of the narrow-band image, with the dotted lines showing the $1\sigma$ of the PSF measurement. These values are very similar to those of the $r^+$ image, see text. }
\label{sizes}
\end{center}
\end{figure*}

As in Nilsson et al. (2007), the sizes of the candidate Ly$\alpha$ and GALEX-detected objects measured in the narrow-band and $r^+$ broad-band images are presented in Fig.~\ref{sizes}. Six LAE objects are excluded due to non-detections in the $r^+$ band. The FWHM was calculated using the FLUX\_RADIUS output from SExtractor, and the PSF in the narrow-band image was calculated by using 9 objects in the image with fluxes in the range of the candidates and the SExtractor parameter CLASS\_STAR larger than 0.8, which inferred a PSF of $0.96''$. The PSF of the $r^+$ image is similar to that of the narrow-band image PSF (the PSF of the $r^+$ image is $1.06''$).
As can be seen in the figure, we have two marginal Ly$\alpha$ blob detections in this survey (of FWHM $3.46''$ and $3.70''$), if a Ly$\alpha$ blob is defined to have a radius larger than $15$~kpc (corresponding to a diameter of $3.75''$). These two (candidates 54 and 66) are also extended in the broad-band images with FWHM $\sim 2''$ and are thus ``normal'' extended Ly$\alpha$ emitters.
We conclude that we have no blobs in this survey. Using the results from Matsuda et al. (2004), Saito et al (2006), Gronwall et al. (2007), and Prescott et al. (2008), the expected number density of Lyman-alpha blobs ranges from a few to a few hundred~$\times 10^{-6}$~Mpc$^{-3}$. The upper limit to the survey is $\sim 3 \times 10^{-6}$~Mpc$^{-3}$, and hence at the low end of the predictions from previous results. Matsuda et al. (2004) and Prescott et al. (2008) argued that Ly$\alpha$ blobs reside in overdense regions of space. If this assumption is correct, it agrees well with the two results from this survey that \emph{i)} we detect a tentatively lower number density of LAEs at this redshift than expected from $z \sim 3$ observations, and \emph{ii)} no Ly$\alpha$ blobs are found in this survey. Both these observations indicate that the COSMOS field is under-dense at $z = 2.25$. We also note that there is a rapid decline in FWHM at larger radii in Fig.~\ref{sizes}, indicating that Ly$\alpha$ blobs and LAEs are two separate categories of Ly$\alpha$ emitting objects. If this was not the case, a tail of objects should be seen towards larger radii. No sample has yet been sufficiently large to confirm whether Ly$\alpha$ blobs and LAEs are separate populations. It is also apparent that the candidate emitters are in general always more extended in the narrow-band image than their broad-band counterparts, i.e. the broad-band counterparts are consistent with being pure point-sources whereas the narrow-band objects are significantly more extended than point-sources (cf. M{\o}ller \& Warren 1998; Fynbo et al. 2001, 2003). This may arise from diffuse scattering of Ly$\alpha$ within the galaxy (see also Laursen \& Sommer-Larsen 2007) and/or due to the diffuse ISM ({\"O}stlin et al. 2008). The extent of this scattering is expected to be much smaller for [OII] emission, hence, the detection of significantly extended emission in a GALEX-detected object may therefore be indicative of it being a Ly$\alpha$ emitter. 

\begin{figure}[!t]
\begin{center}
\epsfig{file=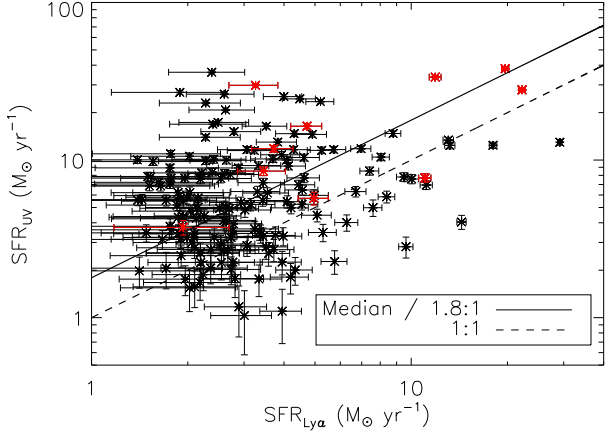,width=9.0cm}
\caption{Star formation rates as derived from the UV rest-frame 1500 {\AA} emission and the Ly$\alpha$ emission for the sample of the 170 candidates. Red points indicate the AGN in the sample, plotted for reference. The measurements correlate well. The solid line indicates the median of the fraction of SFR from the UV and Ly$\alpha$ in the sample, and the dashed line indicates a 1:1 ratio between the two measurements.}
\label{sfr}
\end{center}
\end{figure}

The SFRs of the candidates are calculated using the common star formation indicators of the rest-frame UV continuum emission at $\lambda = 1500$~{\AA} and the Ly$\alpha$ emission. The Ly$\alpha$ emission used in this case was the continuum-subtracted narrow-band flux. The conversion rates that we use are the following:
\begin{equation}\label{eq:sfruv}
\mrm{SFR}_{\mrm{UV}} = \frac{f_{1500\AA} \times 4  \pi d_L^2 \times 1.4 \cdot 10^{-28}}{1+z} \mathrm{M}_{\odot} \mathrm{yr}^{-1}
\end{equation}
where the flux density $f_{1500\AA}$ is in erg~s$^{-1}$~cm$^{-2}$~Hz$^{-1}$, $d_L$ is the luminosity distance in centimeters and $z$ the redshift (Kennicutt 1998). For Ly$\alpha$, the conversion is:
\begin{equation}\label{eq:sfrlya}
\mrm{SFR}_{\mrm{Ly\alpha}} = \frac{F_{\mrm{Ly\alpha}} \times 4 \pi d_L^2}{8.7 \times 1.12 \cdot 10^{41}} \mathrm{M}_{\odot} \mathrm{yr}^{-1}
\end{equation}
where $F_{\mrm{Ly\alpha}}$ is the flux in the Ly$\alpha$ line in erg~s$^{-1}$~cm$^{-2}$ (Brocklehurst, 1971, Kennicutt 1983). A plot of the two resulting SFRs can be seen in Fig.~\ref{sfr}. 
The SFRs are in the range of $\sim 1 - 30$~M$_{\odot}$~yr$^{-1}$ in the Ly$\alpha$ measurement, some of which are unusually high for what is generally seen in Ly$\alpha$ emitters. The dashed line in Fig.~\ref{sfr} corresponds to a 1:1 relation between the UV/Ly$\alpha$ derived SFRs. 
A correlation is seen between the two measurements. The median of the ratio of UV-derived SFR to Ly$\alpha$-derived SFR is 1.8. 
Looking at Eqs.~\ref{eq:sfruv} and \ref{eq:sfrlya}, it is clear that a relation exists between the ratio of the two SFR measurements and the equivalent width of an object, such that:
\begin{equation}\label{eq:sfrratio}
\frac{\mrm{SFR_{UV}}}{\mrm{SFR_{Ly\alpha}}} \propto \frac{cons.}{EW_0}
\end{equation}
In this equation, $EW_0$ is the rest-frame equivalent width of an object. The constant in the nominator depends on the spectral slope of the object, and on dust properties within the galaxy. For a flat slope in $f_{\nu}$ and no dust this constant equals $67$~{\AA}. For this assumption, our median ratio of 1.8 corresponds to a median EW of $37$~{\AA}, which is consistent with the measured median EW of 41.6~{\AA}. 
The ratio is higher than that found by Ouchi et al. (2008), who measured a typical ratio of $\sim 1.2$ in a sample of $z = 3.1$ LAEs, and Gronwall et al. (2007), who, after correcting for the $(1 + z)$ factor, also obtain a typical ratio of $\sim 1.0$ (Gronwall, priv. communication). These results are also related to those presented in Sect.~\ref{ewsec}. We note that several objects have ratios of below one. In total, 19~objects have measurements that deviate by up to $3\sigma$ from the 1:1 line. The explanation for these measurements may originate in the age dependence of the UV SFR conversion factor. According to Kennicutt et al. (1998), the constant in Eq.~\ref{eq:sfruv} is valid for galaxies with constant star formation in the past $\geq 10^8$~yrs. For starbursts of younger ages, the conversion factor should increase. This would bring the ratios closer to the 1:1 line. It appears true for at least 11\% of the sample (19 out of 170), which show signs of young starburst activity. 

\subsection{Equivalent width distribution}\label{ewsec}
\begin{figure*}[!t]
\begin{center}
\epsfig{file=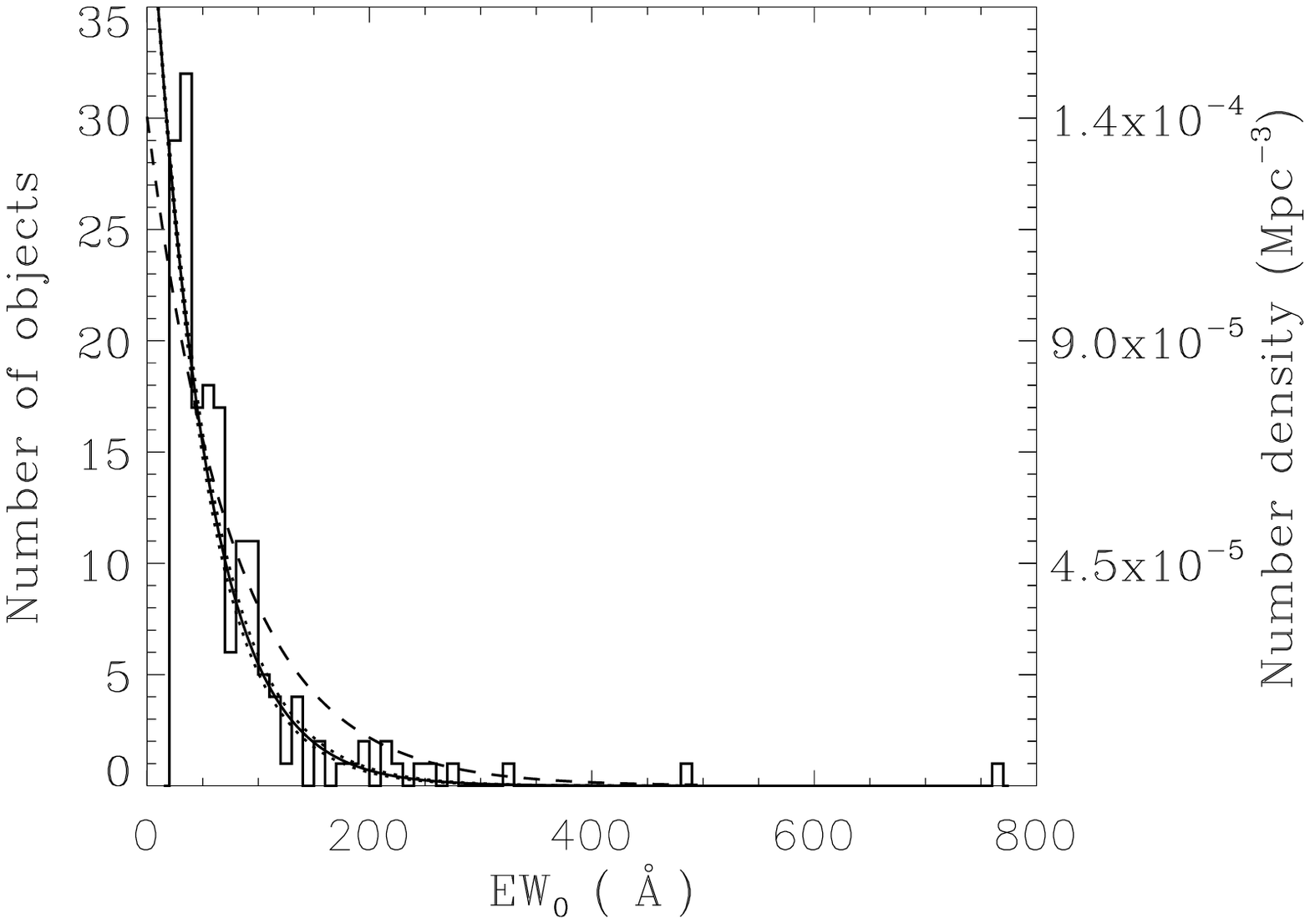,width=9.2cm}\epsfig{file=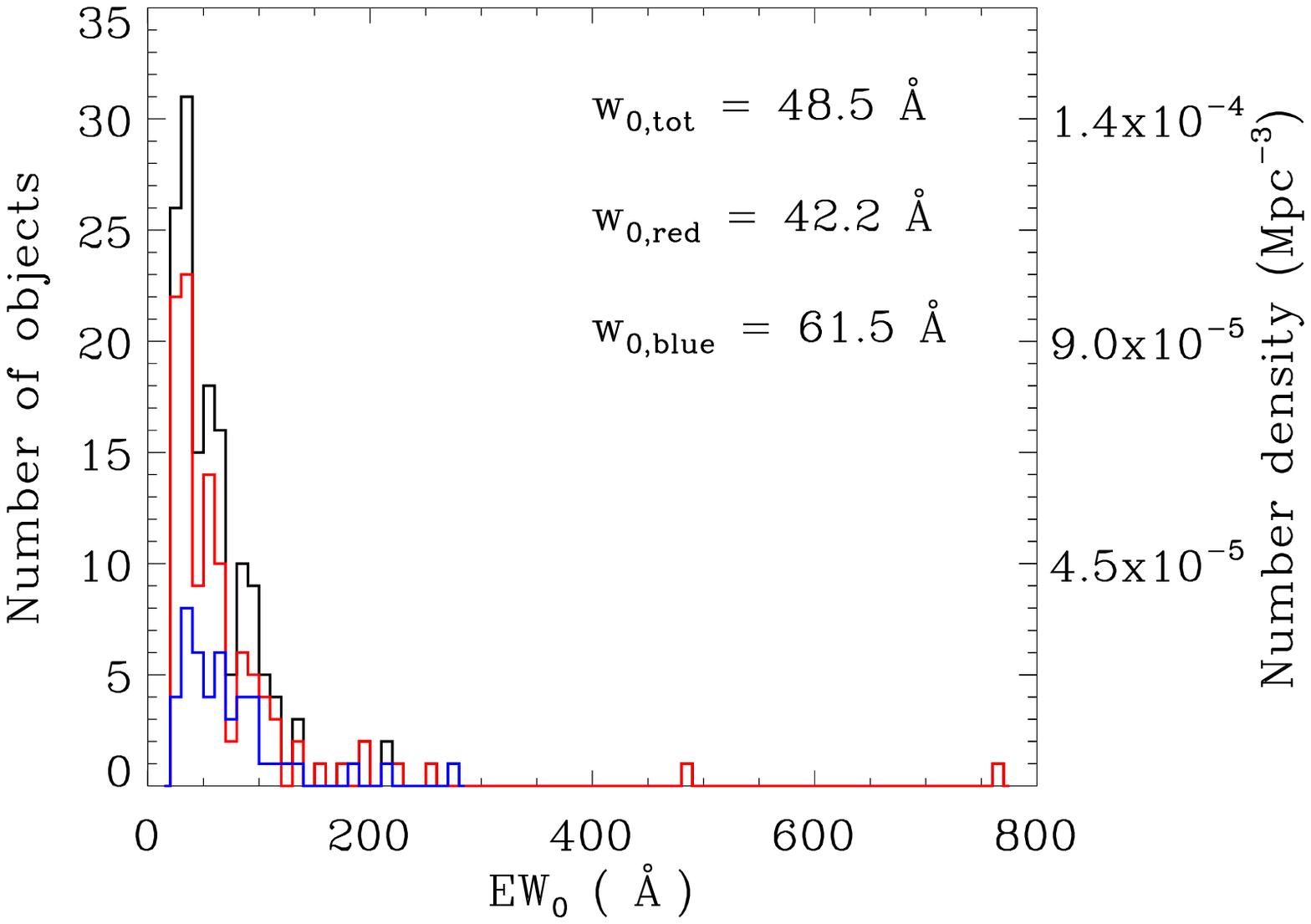,width=9.2cm}
\caption{\emph{Left} Distribution of restframe equivalent widths of the sample of 170 candidates. Bin size is 10~{\AA}. The solid line indicates the exponential fit to the data with the fits with $\pm 1 \sigma$ plotted with dotted lines. The dashed line is the best fit to the sample of Gronwall et al. (2007). This sample has a more narrow distribution of equivalent widths. \emph{Right} Distribution of restframe equivalent widths for total (non-GALEX-detected), red and blue sample (with Vj$ - ^+ >$~or~$< 0$). In this plot, only objects with detections in both the Vj and the $i^+$ bands are given. Red sample is plotted with a red line and blue with a blue line. The width of the different distributions are given in the figure.}
\label{ewdist}
\end{center}
\end{figure*}

At higher redshift ($z \gtrsim 4$), emitters with EWs larger than $240$~{\AA} have been predicted (Schaerer 2003; Tumlison et al. 2003) and observed (Malhotra \& Rhoads 2002; Shimasaku et al. 2006; Stanway et al. 2007). These large EWs supposedly cannot be explained by a normal initial mass function but have been proposed to be signs of Population III stars. At redshift $z \sim 3$, small subsamples of detected emitters exhibit extremely high EWs (a few percent in the surveys of both Gronwall et al. 2007 and Nilsson et al. 2007, and up to 20\% in Ouchi et al. 2008). In Fig.~\ref{ewdist}, the distribution of rest-frame equivalent widths of the candidates can be seen, also divided into the red and blue sub-samples.
The sample includes six emitters with restframe EWs larger than 240~{\AA}, corresponding to $\sim 4$\%. 

In Gronwall et al. (2007), the data is reproduced with an exponentially declining function:
\begin{equation}
N = C \times e^{-\mrm{EW}/w_0}
\end{equation}
where $N$ is the number of objects, $C$ is a normalisation constant and $w_0$ is the e-folding length of the distribution. Gronwall et al. (2007) inferred that $w_0 = 76^{+11}_{-8}$~{\AA}. The EW distribution presented here was fitted here by a best-fit solution minimising the $\chi^2$, and the errors of the fit were calculated by producing $100$~$000$ distributions, normally distributed within the margins of errors in the equivalent widths. Each cumulative distribution was reproduced by an exponential function. No correction for filter transmission curve was done since the filter is nearly box-shaped (see also Fig.~\ref{filtercurve}). The best-fit solution to the e-folding EW for the total non-GALEX-detected sample is $w_{0,tot} = 48.5 \pm 1.7$~{\AA}, less than two thirds of that inferred by Gronwall et al. (2007). This implies that, at $z = 2.25$, there is a higher fraction of LAEs with small EWs than at $z = 3.1$. Corresponding values for the red and blue sub-samples are $w_{0,red} = 42.2 \pm 1.7$ and $w_{0,blue} = 61.5 \pm 4.5$, respectively. The fits indicate that the redder galaxies have a narrower distribution than the bluer galaxies, a result also consistent with the redder galaxies being dustier than the bluer galaxies. This relation was previously observed for LBGs (Shapley et al. 2003).

Since the Ly$\alpha$ EW of a starburst declines with time (Charlot \& Fall 1993; Schaerer 2003) one might seek to interpret the change in width of the distributions in the redshift interval $z =  3 - 2$ as a simple evolution of individual objects. To test this simple idea, we attempted to recreate both the $z = 3.1$ and $z = 2.25$ EW distribution from the simple evolution of single stellar population starbursts by using both the predictions of Charlot \& Fall (1993) and Schaerer (2003). We found that neither of the two distributions can be recreated in this way. The decline in the Ly$\alpha$ EW of a starburst is so rapid that a cumulative EW distribution similar to those in Fig.~\ref{ewdist} will be much narrower than those observed at both redshifts. The only way of obtaining distributions similar to those observed is to allow the majority of the galaxies to have continuous star formation. We can therefore conclude that at maximum a small fraction of the LAEs at redshifts $z = 2 - 3$ are single starburst objects, and are most likely better explained by alternative star formation histories. Finally, the shift in the width of the distribution between $z = 3$ and $z = 2$ cannot be explained by simple evolution with time, but is further evidence of a higher dust content in LAEs at redshift $z \sim 2$. Worth noting is that a narrower EW distribution is consistent with the higher UV-to-Ly$\alpha$ SFRs described in Sect.~\ref{sec:density} since the EW is proportional to the ratio of the flux in the Ly$\alpha$ to the flux density in the UV, similar to the SFRs.

\section{Conclusion}\label{sec:disc}
In this paper, the observations and selection leading to a sample of 170 robust LAE candidates at $z = 2.25$ have been presented. This is the first large-scale survey of LAEs at redshifts around two, and several indications of evolution are seen in the properties of LAEs between redshift $z \sim 3$ and $z \sim 2$.  This is however unsurprising, since the separation in these redshifts corresponds to longer than 1 Gyr. The main arguments for an evolution in the properties of this type of galaxy are:
\begin{itemize}
\item[-] \emph{The large spread in SEDs}
\end{itemize} 
The most conspicuous evolution detected was that of the emergence of a large number of ``red'' LAEs. The 12 individual detections in the relatively shallow $K_s$ images and the red mean SED of the total stacked sample, presented in Fig.~\ref{fig:redsed}, indicated that a significant subsample, of up to two-thirds of the total sample, of LAEs are massive, and potentially dusty, galaxies. We also showed that the rest-frame UV colours of these galaxies cannot be explained by a simple stellar population, and that even though the majority of the candidates are consistent with being young galaxies, a significant fraction of the sample displays more complicated colours. 
\begin{itemize}
\item[-] \emph{Potentially larger fraction of AGN}
\end{itemize}
The result above is also related to the detection of at least nine AGN in the sample. Again, this is not a surprising result since redshift $z \sim 2$ corresponds to the peak of the AGN number density distribution (see e.g. Wolf et al. 2003 and references therein). The number density of AGN in this sample is in principle consistent with surveys at higher redshift, but is likely to be higher by a factor of two.
\begin{itemize}
\item[-] \emph{Change in SFR ratios}
\end{itemize}
The ratio of UV to Ly$\alpha$ derived SFRs has a higher median value in this survey than results at redshift $z \sim 3$. The spread is large, but the trend is clear. This result, and the smaller EWs discovered in general for this sample (see also the next point) at high redshift may indicate that LAEs in this survey are expected to be more affected by dust than LAEs at $z \sim 3$. 
\begin{itemize}
\item[-] \emph{Narrower EW distribution}
\end{itemize}
Another intriguing and related result is that of the narrower equivalent width distribution. Firstly, as detailed in section~\ref{ewsec}, the distribution is difficult to explain without invoking complicated arguments about the properties of the galaxies. A possible explanation would be that the star formation histories of Ly$\alpha$ emitters are complex. Secondly, the difference in the distributions between redshifts three and two is further evidence of a higher dust quantity in the lower redshift galaxies.

In conclusion, by comparing observations of Ly$\alpha$ emitters at redshift $z = 2.25$ with galaxies selected in the same manner at higher redshifts, several evolutionary signatures become evident in the properties of these galaxies. At lower redshifts, there appear to be fewer objects, with redder colours and higher dust contents, smaller equivalent widths, and a higher fraction of objects containing AGN. Future SED fitting of these galaxies will reveal more information into the properties such as dust, mass and age (Nilsson et al., in prep.).

\begin{acknowledgements}
The Dark Cosmology Centre is funded by the DNRF. The authors wish to thank Hermann-Josef R{\"o}ser for kind help with \emph{MPIA}phot and Katherine Inskip and Joanna Holt for help with the COSMOS HST images. 
\end{acknowledgements}

\end{document}